\documentclass[twocolumn,showpacs,superscriptaddress,preprintnumbers,amsmath,amssymb,prb]{revtex4-1}
\usepackage{amsmath,amsfonts,bm,graphicx}
\usepackage{color}

\newcommand{\expect}[1]{\langle #1\rangle}

\begin{document}

\title{
First-Principles Electron Transport with Phonon Coupling: Large-Scale at Low Cost
}

\author{Tue Gunst}
\email{Tue.Gunst@nanotech.dtu.dk}
\affiliation{Department of Micro- and Nanotechnology (DTU Nanotech), Center for Nanostructured Graphene (CNG), Technical University of Denmark, DK-2800 Kgs. Lyngby, Denmark}
\author{Troels Markussen}
\affiliation{QuantumWise A/S, Fruebjergvej 3, Postbox 4, DK-2100 Copenhagen, Denmark}
\author{Mattias L. N. Palsgaard}
\affiliation{QuantumWise A/S, Fruebjergvej 3, Postbox 4, DK-2100 Copenhagen, Denmark}
\author{Kurt Stokbro}
\affiliation{QuantumWise A/S, Fruebjergvej 3, Postbox 4, DK-2100 Copenhagen, Denmark}
\author{Mads Brandbyge}
\affiliation{Department of Micro- and Nanotechnology (DTU Nanotech), Center for Nanostructured Graphene (CNG), Technical University of Denmark, DK-2800 Kgs. Lyngby, Denmark}

\date{\today}
\begin{abstract}
Phonon-assisted tunneling plays a crucial role for electronic device performance and even more so with future size down-scaling. We show how one can include this effect in large-scale first-principles calculations using a single "special thermal displacement" (STD) of the atomic coordinates at almost the same cost as elastic transport calculations, by extending the recent method of Zacharias et al. [Phys Rev. B 94, 075125 (2016)] to the important case of Landauer conductance.
We apply the method to ultra-scaled silicon devices and demonstrate the importance of phonon-assisted band-to-band and source-to-drain tunneling.
In a diode the phonons lead to a rectification ratio suppression in good agreement with experiments, 
while in an ultra-thin body transistor the phonons increase off-currents by four orders of magnitude, and the subthreshold swing by a factor of four, in agreement with perturbation theory. 
\end{abstract}


\maketitle

Electron-phonon inelastic scattering is one of the major challenges for emerging high-performance ultra-scaled devices, from the viewpoint of both experiments and device simulations\cite{ionescu_tunnel_2011}. 
Semi-classical device simulations fail to describe quantum tunneling while atomistic quantum simulations often are too time-consuming to treat phonon scattering accurately.
Reducing the computational cost of inelastic, compared to elastic, device simulations has therefore been an important and unsolved challenge for decades since the first ultra-scaled transistors emerged.
In the extreme limit of molecular-scale devices there are accurate first-principles methods for inelastic processes available\cite{lu_efficient_2014,frederiksen_inelastic_2007,lee_efficient_2016,mera_hypergeometric_2016,lee_quantum_2017,vandenberghe_generalized_2011,niquet_quantum_2014,jiang_first-principles_2005,troisi_molecular_2006,galperin_molecular_2007,gustafsson_theory_2014,troisi_vibronic_2003}, while in the  opposite bulk continuum limit, deformation potentials (DPs) are extracted for Boltzmann transport equations (BTEs) that accurately describe low bias transport\cite{gunst_first-principles_2016,hwang_acoustic_2008,giustino_electron-phonon_2017,fischetti_band_1996}.
However, in between these two regimes efficient computational methods are missing. 
One approach is to apply the continuum DP, despite the fact that electron-phonon coupling (EPC) is known to change significantly in nanostructured devices\cite{bozyigit_soft_2016,zhang_atomistic_2010,milnikov_random_2016} and in an electrostatic environment\cite{gunst_flexural-phonon_2017}.
Alternatively, it is possible to perform atomistic tight-binding calculations with coarse diagonal self-energy approximations at an extensive computational cost\cite{luisier_atomistic_2009,rhyner_atomistic_2014}.
Modern computers are unable to include EPC from first-principles beyond the molecular scale, while
the understanding and design of emerging ultra-scaled devices calls for atomistic simulations with an accurate description of EPC for thousands of atoms including quantum confinement, strain and surface effects.

Stochastic sampling of lattice fluctuations, through molecular dynamics\cite{markussen_electron-phonon_2017,liu_direct_2015,li_conduction_2004,andrews_stochastic_2008,paulsson_conductance_2009,solomon_quantum_2008,dreher_structure_2005} (MD) and Monte Carlo\cite{zacharias_stochastic_2015,monserrat_vibrational_2016,pecchia_role_2003}, has previously been used to estimate the variation of the Landauer conductance or dielectric function with temperature.
Key motivations in these developments are the conceptual simplicity and computer memory efficiency compared to perturbation theory (PT).
The MD is able to capture anharmonic effects, but is limited to the classical high temperature regime for systems with light atoms, neglecting zero-point motion and low temperature freeze-out of phonons\cite{pecchia_role_2003,zacharias_stochastic_2015,ponce_temperature_2015,monserrat_vibrational_2016}. 
However, the computational cost of sampling all atomic displacements in the configuration space, introduce yet a system-size-scaling cost which remain an obstacle in all these methods.

Recently, Zacharias \textit{et al.}\cite{zacharias_one-shot_2016} showed that the stochastic sampling of configurations can be replaced by a {\em single} optimal supercell configuration for band gap renormalization and phonon-assisted optical absorption.
Inspired by the work of Zacharias \textit{et al.}, we present in this letter a "special thermal displacement" (STD) method based on nonequilibrium Green's functions (NEGF). 
The STD method is able to deterministically handle EPC in systems with thousands of atoms with a computational burden equivalent to that of elastic transport.
This extends the capability of computer simulations to handle nm-scaled devices.
The method applies to systems with a high degree of repetition of the same basic unit cell since it relies on cancellations of errors between degenerate phonon modes.
Often good force-fields exist in such systems while the electron-phonon coupling is less well described. We therefore combine phonons obtained by a force-field with the EPC evaluated from Density Functional Theory (DFT).
We target systems which have a bulk-like representation of vibrations (non-localized) which is the case for a large selection of technologically important devices. 
As key examples, we study the properties of bulk silicon, the performance of silicon based rectifiers, and double-gated metal-oxide-semiconductor field-effect transistors (MOSFETs).
We demonstrate how EPC can be studied by first-principles calculations for systems with thousands of atoms using modest computer resources, while yielding results consistent with PT for smaller systems. This makes the STD method a promising nanoscale design tool for predicting trends in realistic nano-devices under working conditions.

{\it Finite temperature phonon-assisted tunneling.}
The starting point is to consider the adiabatic limit of slowly moving atoms where we consider the parametric dependence of the retarded device Green's function, $\mathbf{G}^r(E,\{\mathbf{u}_{\lambda}\})$, on the nuclear displacements, $\mathbf{u}_{\lambda}(T,V)$. 
The thermally averaged current is given by,\cite{SupplementalMaterial}
\begin{eqnarray}
I(V,T) &=&  \frac{2e}{h} \int dE \expect{\mathcal{T}(E,T)} \left[f_L-f_R\right]
\,,\nonumber\\
\expect{\mathcal{T}(E,T)} &=& \Pi_{\lambda} \int du_{\lambda} \frac{\text{exp}\left(-u_{\lambda}^{2}/2 \sigma_{\lambda}^{2}\right)}{\sqrt{2 \pi} \sigma_{\lambda}} \mathcal{T}(E,\{\mathbf{u}_{\lambda}\})\label{eqn:TransmissionTdependent}
\end{eqnarray}
where $\mathcal{T}(E,\{\mathbf{u}_{\lambda}\}) = Tr\left[\mathbf{\Gamma}_L \mathbf{G}^r(\{\mathbf{u}_{\lambda}\}) \mathbf{\Gamma}_R \mathbf{G}^a(\{\mathbf{u}_{\lambda}\})\right]$, $\mathbf{\Gamma}_{\alpha}$ are the electrode coupling matrices,
 and $f_{\alpha}$ the Fermi-function at the chemical potential of lead $\alpha$.
The phonon modes are labeled by $\lambda$ with frequency $\omega_{\lambda}$, eigenmode vector $\mathbf{e}_{\lambda}$, and characteristic length, $l_{\lambda}$. The Gaussian width $\sigma$ is related to the mean 
square displacement $\langle \mathbf{u}_{\lambda}^{2} \rangle = l^{2}_{\lambda} (2 n_B(T) +1) = \sigma_{\lambda}^{2}(T)$ at a temperature $T$. 
In principle these integrals can be computed directly for small systems by Gaussian quadratures or by Monte Carlo importance sampling to obtain the average over the ensemble of possible atomic positions.
However, a single STD, $\mathbf{u}_{STD}$, is sufficient for large systems with a high repetition of smaller unit cells, defined as
\begin{eqnarray}
\mathbf{u}_{STD}(T) = \sum_{\lambda} s_{\lambda} (-1)^{\lambda-1} \sigma_{\lambda}(T) \mathbf{e}_{\lambda}\label{eqn:OSdisplacement}
\end{eqnarray}
Here $s_{\lambda}$ denotes the sign of the first non-zero element in $\mathbf{e}_{\lambda}$ enforcing the same choice of "gauge" for the modes.
Our equations are in the form similar to the dielectric function of bulk systems considered by Zacharias \textit{et al.\cite{zacharias_one-shot_2016,STDname}}. 
For completeness we repeat the argument\cite{zacharias_one-shot_2016} stating that the STD configuration gives the correct thermal average for large systems by comparing the Taylor expansion of Eq.~\ref{eqn:TransmissionTdependent} around the equilibrium configuration evaluated at the mode-displacements, 
\begin{eqnarray}
\mathcal{T}(E,\{\mathbf{u}_{\lambda}\}) 
&=& \mathcal{T}_0(E) + \sum_{\lambda}\frac{\partial\mathcal{T}(E,\{\mathbf{u}_{\lambda}\})}{\partial \mathbf{u}_{\lambda}}\mathbf{u}_{\lambda}\nonumber
\\&+&\frac{1}{2} \sum_{\lambda} \frac{\partial^2\mathcal{T}(E,\{\mathbf{u}_{\lambda}\})}{\partial \mathbf{u}_{\lambda}^{2}}\mathbf{u}_{\lambda}^{2}+\mathcal{O}(\sigma^{3})\,,
\label{eqn:DirectTaylor}
\\
\expect{\mathcal{T}(E,T)}
&=& \mathcal{T}_0(E) + \frac{1}{2} \sum_{\lambda} \sigma_{\lambda}^{2}(T) \frac{\partial^2\mathcal{T}(E,\{\mathbf{u}_{\lambda}\})}{\partial \mathbf{u}_{\lambda}^{2}} + \mathcal{O}(\sigma^{4}) \nonumber
\end{eqnarray}
to the Taylor expansion around the STD configuration $\mathbf{u}_{STD}$ evaluated at zero:
\begin{eqnarray}
&&\mathcal{T}_{STD}(E,T) 
= 
\mathcal{T}_0(E) - \sum_{\lambda}\frac{\partial\mathcal{T}(E,\{\mathbf{u}_{\lambda}\})}{\partial \mathbf{u}_{\lambda}} s_{\lambda} (-1)^{\lambda-1} \sigma_{\lambda}(T)
\label{eqn:OStaylor}\\
&&+ \frac{1}{2} \sum_{\lambda\lambda'} \frac{\partial^2\mathcal{T}(E,\{\mathbf{u}_{\lambda}\})}{\partial \mathbf{u}_{\lambda}\partial \mathbf{u}_{\lambda'}}
s_{\lambda}s_{\lambda'} (-1)^{\lambda+\lambda'-2} \sigma_{\lambda}(T)\sigma_{\lambda'}(T)
+ \mathcal{O}(\sigma^{3})\nonumber
\end{eqnarray}
The two successive terms, in the sum of the first order part of Eq.~\ref{eqn:OStaylor}, cancel each other since for large systems the two phonon modes $\lambda$ and $\lambda+1$ are near degenerate resulting in an equivalent electron-phonon coupling and transmission derivatives.
The second order term in Eq.~\ref{eqn:OStaylor} is finite only for $\lambda=\lambda'$ and specifically $\lambda$ and $\lambda+1$ terms once again have opposite signs.
Hereby the STD expression Eq.~\ref{eqn:OStaylor} approaches the direct result Eq.~\ref{eqn:DirectTaylor} for $N\rightarrow \infty$.
According to Ref.~\onlinecite{zacharias_one-shot_2016} the accuracy can be controlled not only by system size but also by configurational averaging over configurations with a systematically flipped sign in a subset of the mode displacements in Eq.~\ref{eqn:OSdisplacement}.
Unlike PT, which relies on a series truncation at the lowest $\mathcal{O}(\sigma^{2}) \sim \mathcal{O}(u^2)$, the STD expression holds to all orders in $\sigma_{\lambda}$. This is consistent with the adiabatic assumption of large displacements and low velocities. 
The current in Eq.~\ref{eqn:TransmissionTdependent} evaluated from the STD, Eq.~\ref{eqn:OSdisplacement}, provides a simple model treating phonon-assisted tunneling and temperature dependent EPC renormalization of the electronic structure on an equal footing.
The STD approximates the correct thermal average, $\expect{\mathcal{T}(E,T)} \approx \mathcal{T}(E,\mathbf{u}_{STD}(T))$, of the Landauer conductance, and resembles the special quasi-random structures (SQS) used to model infinite random alloys\cite{wei_electronic_1990}. 
The phonon occupations could include a contribution, in addition to the thermal $n_B$, from finite bias heating. This would pave the way for current-saturation and heating modeling in nanoscale devices in the future.

\begin{figure}[!htbp]
\centering
{\includegraphics[width=0.99\linewidth]{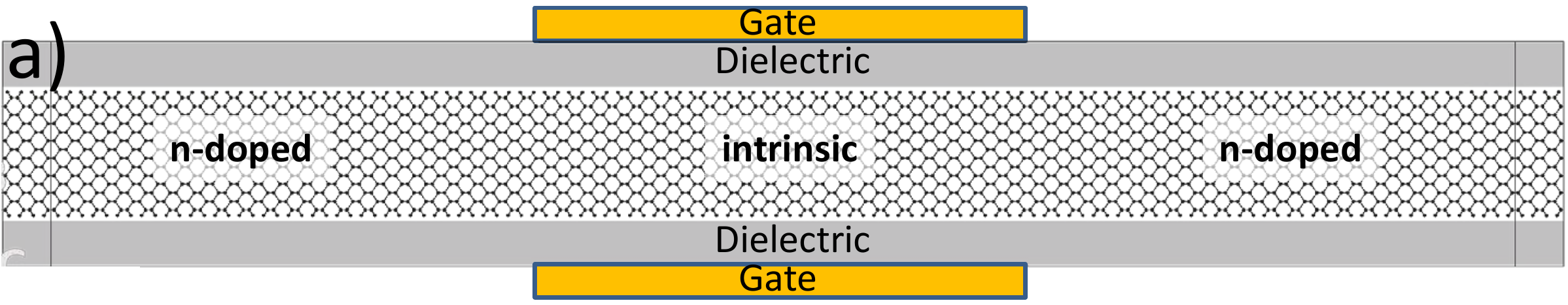}}\\
{\includegraphics[width=0.99\linewidth]{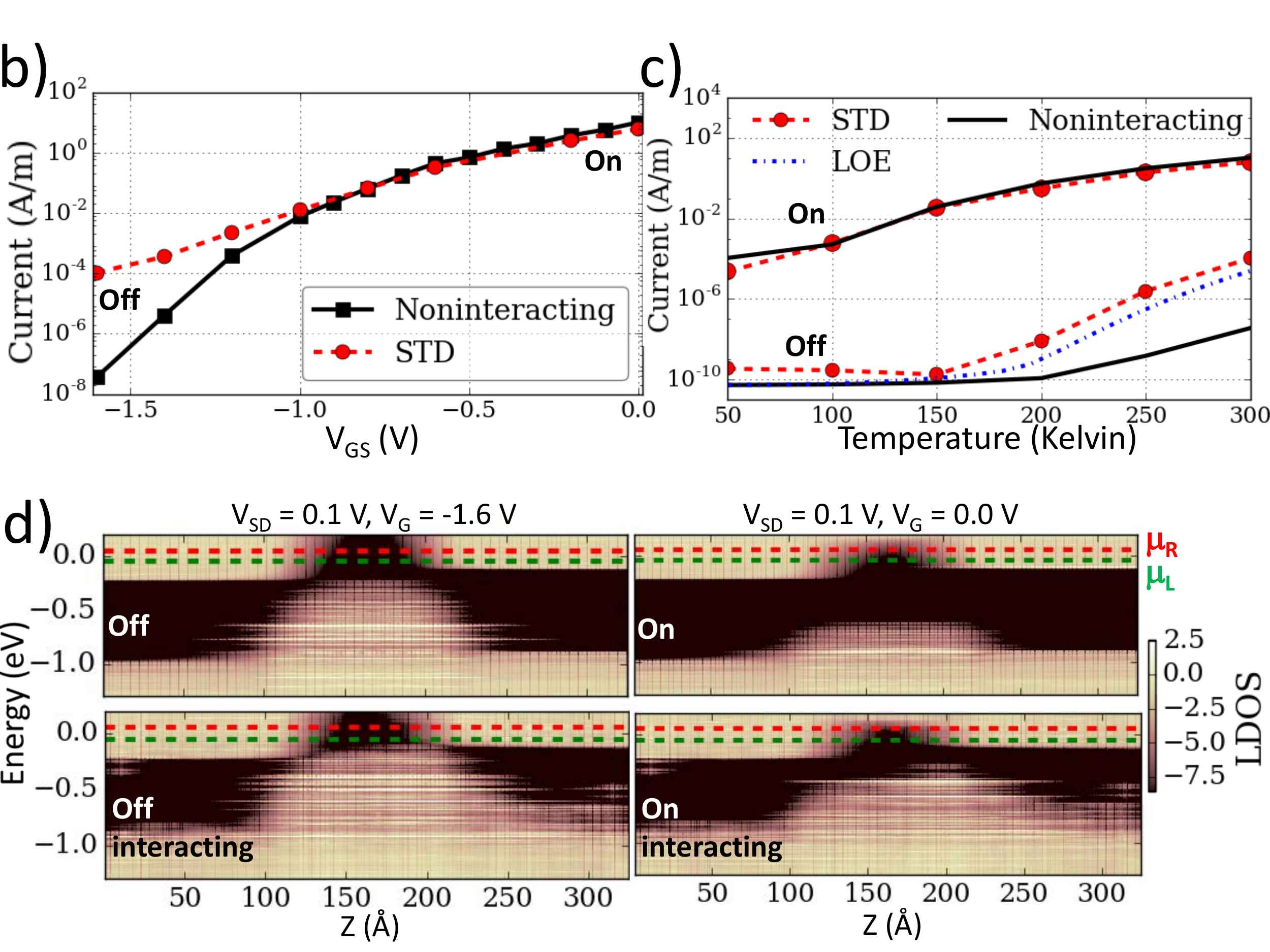}}
\caption{
(a) Silicon $n$-$i$-$n$ junction tunneling device with a source and drain doping of $1.0\times 10^{21}$\,cm$^{-3}$, length of 33\,nm and $\sim\,2000$ atoms.
(b) Current versus gate-voltage, $V_{G}$, for a source-drain voltage $V_{SD}$=0.1\,V and at 300\,K.
(c) Current at $V_{G}$=-1.6\,V (on) and $V_{G}$=0\,V (off) as a function of temperature for $V_{SD}$=0.1\,V.
(d) Tunneling barrier dependence on $V_{G}$ and temperature illustrated by the local device density of states (LDOS). Regions of low (high) LDOS are shown in dark (bright) color illustrating the forbidden (permitted) bands along the device. Finite temperature electron-phonon coupling increases the off-state current hereby degrading the device performance.
}
\label{fig:NINresults}
\end{figure}

{\it Silicon $n$-$i$-$n$ junction device.}
We now turn to device characteristics including EPC\cite{ATK}. 
Figure~\ref{fig:NINresults} presents full quantum device simulations including EPC for a two-dimensional Si $n$-$i$-$n$ double-gated MOSFET with 10\,nm gate length.
Decreasing the gate-voltage the device goes from an on-state where the current originates from thermionic emission to an off-state where the current is determined by source-to-drain tunneling through the barrier. 
Comparing the interacting STD-Landauer result with the elastic calculation shows that the on-current is almost unchanged by phonon scattering, Fig.~\ref{fig:NINresults}b. The on-current reaches a value of $\sim 10$\,A/m even with phonon-scattering at 300\,K.
However, phonon-assisted tunneling is found to increase the off-state current by four orders of magnitude.
Consequently, we extract a significant subthreshold swing ($S$) degradation from $S \approx 97$\,mV/dec to $S \approx 375$\,mV/dec at 300\,K.
Existing device simulations on silicon FETs have not reported any significant phonon-assisted $S$ degradation, most likely because they either neglect quantum-tunneling, or are based on deformations potentials (corresponding to a purely imaginary and diagonal self-energy in the NEGF formalism) and effective-mass or tight-binding approximations\cite{cavassilas_one-shot_2013,mori_effects_2008,svizhenko_role_2003,koswatta_possibility_2010}.
A single study found a significant increase in the subthreshold current in SiNWs partly traced back to the renormalization (self-energy real-part), however still within deformation potential approximations\cite{valin_quantum_2014}.
In Fig.~\ref{fig:NINresults}d we illustrate the temperature dependent broadening and shift of the density of states that effectively modifies the barrier thickness and phonon-assisted tunneling rates from electron states with $s/d$-type orbital character through evanescent $p$-type states 
in the intrinsic barrier region. Since elastic tunneling is suppressed by the orbital symmetry, we find that the off-current is highly sensitive to temperature and significantly increased by EPC at finite temperature.

These results agree with quantum PT, as implemented in the lowest order expansion (LOE) method\cite{lu_efficient_2014,gunst_inelastic_2016}. The LOE calculation essentially requires evaluation of the transition rates between scattering states for each phonon mode one-by-one.
This makes a full LOE calculation computationally more expensive by a factor of at least 6000 from the number of phonon modes present in the device. This is a tremendous task and to achieve this for a single gate-value we employ several computational approximations\cite{SupplementalMaterial}.
In Fig.~\ref{fig:NINresults}c, we show the temperature dependence of the on- and off-currents and validate the STD-Landauer result with the computational expensive LOE calculation for the off-state. Importantly, we obtain an excellent match between the LOE and STD-Landauer method.
The temperature dependence of the current shows that phonon-assisted tunneling is frozen-out below 150\,K. Similarly, other simulations have found that phonon broadening of single impurity levels in SiNWs suppress current saturation above 150\,K\cite{bescond_size_2016}. 
In conclusion, phonon-assisted tunneling is found to play a major role for leakage currents in ultra-thin body transistors at room temperature.

\begin{figure}[!htbp]
\centering
{\includegraphics[width=0.99\linewidth]{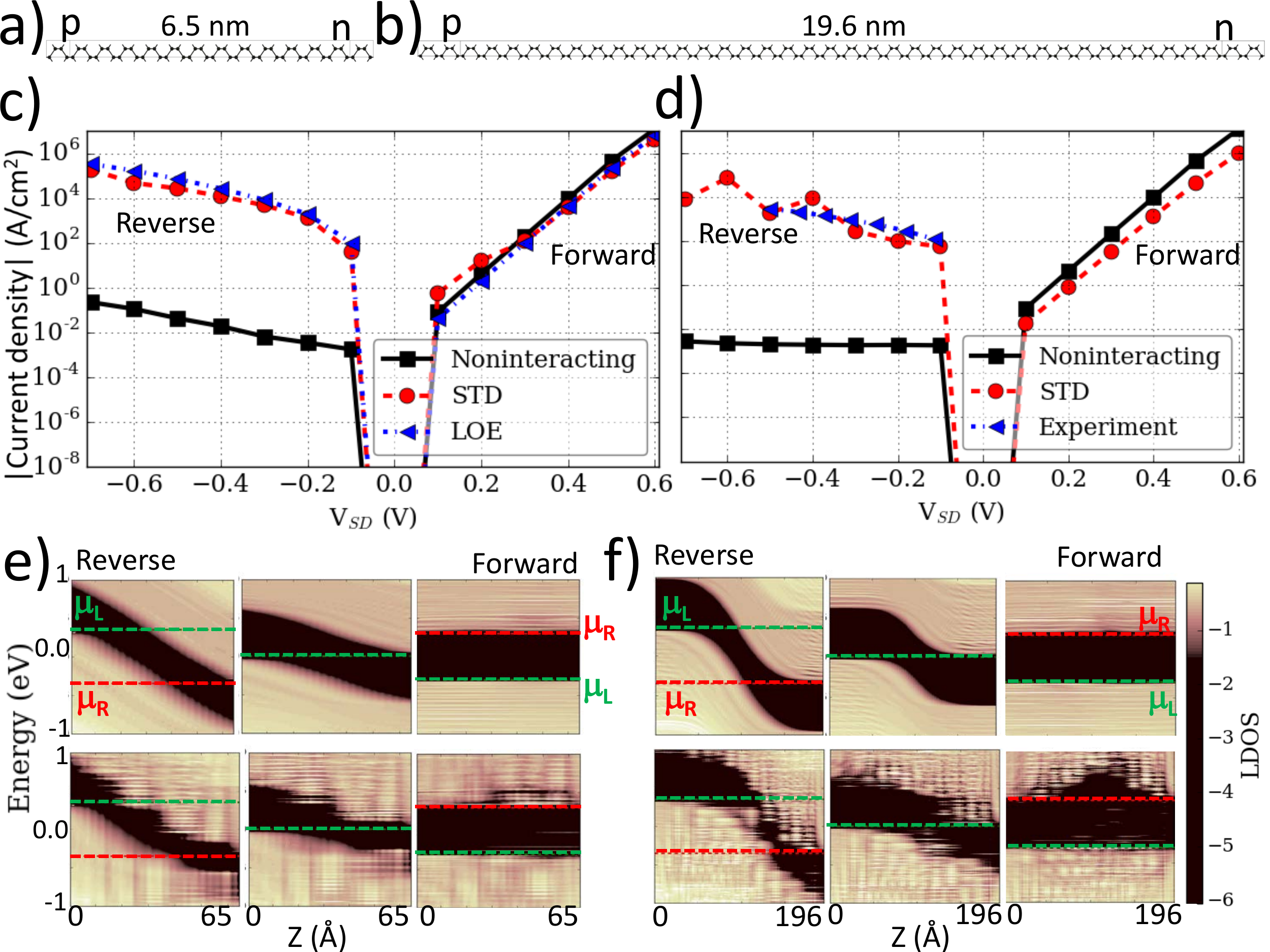}}
\caption{Silicon $p$-$n$ junction devices with doping values of $\pm 2\times 10^{19}$\,cm$^{-3}$. Device characteristics are compared between a short (a,c) and long (b,d) rectifier. The short device permits comparison with perturbation theory (LOE) while the screening is not fully enclosed within the device as shown by the local density of states (e). In both short and long devices the LDOS is strongly renormalized at 300\,K (e,f lower panels) and the reverse leakage currents are increased by six orders of magnitude. Experimental off-current in (d) from Ref.~\onlinecite{schmid_silicon_2012} for a slightly asymmetrically doped SiNW at the Esaki onset.}
\label{fig:PNresults}
\end{figure}

{\it Silicon Rectifiers.}
Next we show that finite temperature EPC does not only increase source-to-drain tunneling, but also significantly increases the band-to-band tunneling in $p$-$n$ junctions.
In Fig.~\ref{fig:PNresults} we consider transport in a short (6.5~nm) and a long (19.6~nm) silicon $p$-$n$ junction\cite{schmid_silicon_2012,vandenberghe_generalized_2011,rhyner_atomistic_2012} with transport in the $[100]$ crystal direction.

Figure~\ref{fig:PNresults}c(d) shows the modification of the IV-characteristics due to EPC in the short (long) rectifier.
To demonstrate the validity of the STD-Landauer method, we start by comparing the IV curves obtained with that from a PT(LOE) calculation\cite{lu_efficient_2014,gunst_inelastic_2016}. Again, the PT calculation is computationally more expensive by a factor of at least 150 from the number of modes in the device. Nevertheless, we obtain an almost perfect match between the two in Fig.~\ref{fig:PNresults}c.

One challenge for DFT simulations of silicon devices is the fact that the screening length is often longer than system sizes reachable by PT calculations.
This is illustrated by the local density of states (LDOS) in Fig.~\ref{fig:PNresults}e-f which show how the typical $p$-$n$ junction potential profile emerges when increasing the device length. 
As shown in Fig.~\ref{fig:PNresults}f, the STD-Landauer approach enables large-scale device simulations including EPC that secures converged screening potentials.
In addition, we also see that EPC gives rise to significant changes in the LDOS of the device that highlights the importance of EPC in device characterization.
Device performance is measured by its ability to have a high forward current, $I_{ON}$, and a low reverse leakage current, $I_{OFF}$.
The $I_{ON}$/$I_{OFF}$ figure-of-merit is reduced from $2\times10^8$ to $4$ at $\pm0.5$~V and $6\times10^9$ to $5\times10^2$ at $\pm0.6$~V due to EPC.
The reverse current still saturates, but at a much higher value. Hereby the low bias performance in terms of the rectification ratio is ruined demonstrating how the EPC can have detrimental impact on the rectification ratio
 and consequently a higher power is needed for efficient rectification. 

The STD-Landauer result shows an increasing off-current due to phonon excitation when increasing the temperature to 300\,K.
Recent experiments performed by Schmid \textit{et al.}\cite{schmid_silicon_2012} on $pn$-junctions made from silicon nanowires with a diameter of 60\,nm report on several key features that match our findings.
Their experiments at different temperatures underlines the pivotal role played by phonons in the device characteristics. They explore a range of dopings going from normal to Esaki diode characteristics.
At room temperature and at the lowest doping corresponding to the onset of Esaki characteristics, they find a maximum off-current density of $10^3$A/cm$^2$ at a reverse bias of -0.5\,V.
Our device is at a doping level just before the onset of Esaki characteristics, where Fermi-levels are still inside the gap, cf. Fig.~\ref{fig:PNresults}f. The doping onset of the Esaki regime serves as a good point of reference since it is independent of the band gap value.
In agreement with the experiments we estimate $I_{OFF}(-0.5\,V) \approx 10^3$A/cm$^2$ and also find $I_{ON}$/$I_{OFF}<1$ below $\pm0.5$\,V, while the noninteracting ballistic result is off by roughly six orders of magnitude.
In addition, the experiment shows a strong temperature dependence of the off-current indicating an increased probability for transmission across the junction consistent with the additional transport channels opened by EPC in our simulations.
Unlike the ballistic noninteracting case we find that $I_{OFF}$ increases with bias, Fig.~\ref{fig:PNresults}d. This is traced back to an increased window for inelastic transmission across the device that scales with the bias window.
Again, this trend fits with the experiments performed by Schmid \textit{et al.}\cite{schmid_silicon_2012}

{\it Carrier mobilities.}
Carrier mobilities limited by EPC is an important performance indicator of materials.
Finally we show that the STD-Landauer approach has a predictive power at the level of state-of-the-art BTE solvers\cite{gunst_first-principles_2016} based on the full first-principles EPC, and that both methods are in excellent agreement with available experimental results.
\begin{figure}[!htbp]
\centering
{\includegraphics[width=0.99\linewidth]{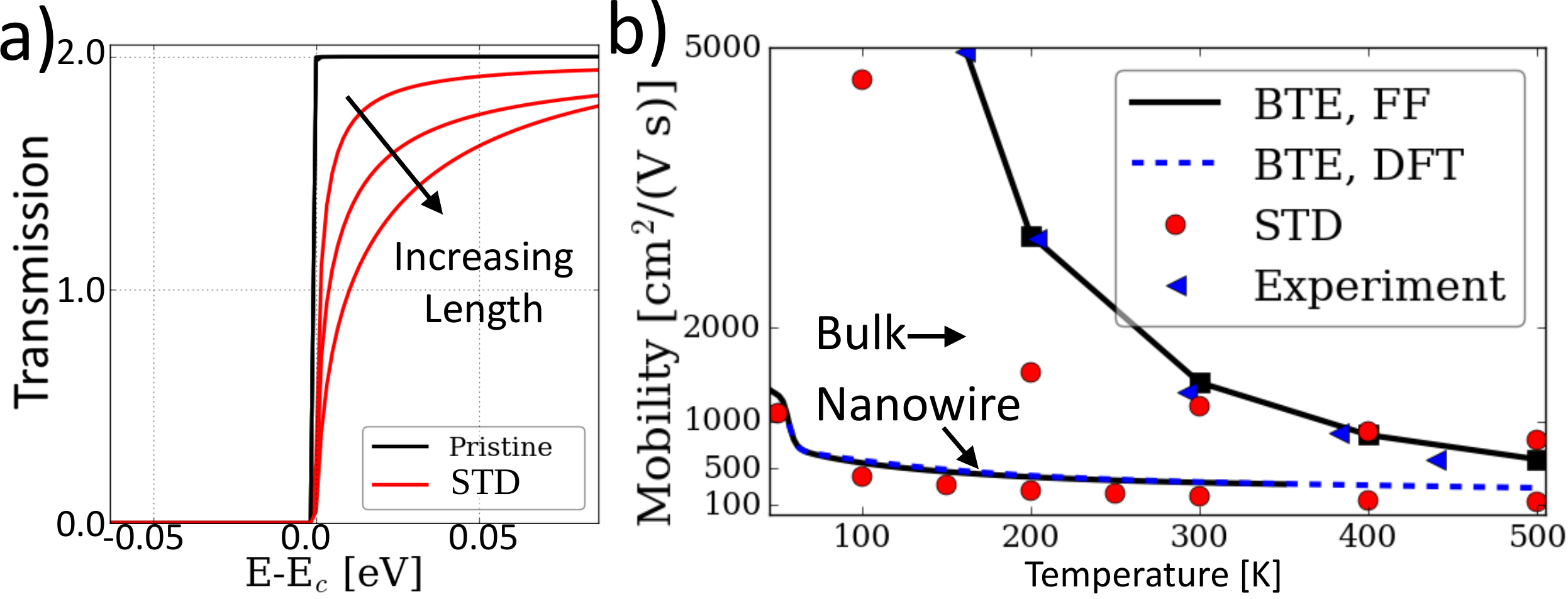}}
\caption{STD-Landauer mobilities. (a) Transmission for the pristine and STD configurations for three different lengths of a 1.3\,nm diameter silicon nanowire. (b) Mobilities of a silicon nanowire and bulk silicon. STD result is compared to BTE as well as experimental data from Ref.~\onlinecite{jacoboni_review_1977} and force-field (FF) phonons for bulk and nanowire, respectively.}
\label{fig:mobility}
\end{figure}
In Fig.~\ref{fig:mobility}, we present mobilities obtained from the STD-Landauer device model.
The resistance $R(\mathcal{L},T) = 1/G(\mathcal{L},T) = R_c + \rho_{1D}(T) \mathcal{L}$ increases linearly with length, $\mathcal{L}$, of the dynamic region in the ohmic regime.
In Fig.~\ref{fig:mobility}a we show the transmissions at 300\,K for increasing device lengths. From this we extract a one-dimensional resistivity, $\rho_{1D}(T)$, which depends on temperature but not on wire length, and the contact resistance, $R_c$.
From the density of states, $D(E)$, and carrier density, $\tilde{n}= \int_{E_g}^\infty f(E-E_F,T) D(E)dE$, we can obtain a mobility $\mu = \frac{1}{q\,\tilde{n}\,\rho_{1D}}$. The obtained values for bulk silicon compares well with both experimental values as well as BTE results from room temperature.
The STD-Landauer result includes multi-phonon effects and assumes the correct quantum occupations where optical modes are frozen-out at low temperatures. The adiabatic assumption neglects, however, the frequency in single-phonon emission for modes with a high frequency which may explain part of the discrepancy at low temperature.
Our first-principles calculations further support the conclusion of enhanced electron-phonon coupling in nanowires\cite{zhang_atomistic_2010,markussen_electron-phonon_2017,FN_MFP}.
In addition, we compare the results obtained with both force-field and DFT phonons for the SiNW giving almost the exact same values. The predictability of the STD-Landauer approach does in general not rely on an accurate description of a single phonon mode but rather the full configuration space. Hereby force-fields become even more relevant for device simulations.

{\it Conclusions.}
We have presented how a single “special thermal displacement” (STD) together with a Landauer conductance calculation enables nanometer-scale nonequilibrium device simulations including phonon-assisted tunneling and temperature renormalization from first-principles.
Our results are in excellent agreement with both experiments and state-of-the-art perturbation theory calculations and underlines the key role played by phonon-assisted band-to-band and source-to-drain tunneling in the performance of ultra-scaled silicon rectifiers and transistors.
Tunneling from electron states with $s/d$-character through evanescent $p$-type states in the transistor barrier may put a limit to the performance of sub-10-nm devices and the length-scale where elastic and classical device simulations are reliable.
Importantly, the STD-Landauer approach is far more memory and computational efficient making it appealing as an atomistic design tool in electronics.
The STD method evaluates phonon coupling under operating conditions and in the future it may open up the possibility for efficient modeling of current-induced heating by letting the phonon occupations depend on the applied bias voltage.

%

\begin{acknowledgments}
The authors acknowledge support from Innovation Fund Denmark through Grant No. 79-2013-1 and the Quantum Innovation
Center (QUBIZ). CNG is sponsored by the Danish National Research Foundation, project No. DNRF103.\nocite{brandbyge_density-functional_2002,stradi_general_2016,tersoff_empirical_1988,Feynmann_statistical_physics,patrick_unified_2014}
\end{acknowledgments}


\bibliography{EPCinLargeSystemsPaper}

\begin{thebibliography}{55}
\bibitem [{FN_()}]{FN_MFP}%
  \BibitemOpen
  \href@noop {} {\bibinfo  {journal} {One can also extract the mean-free-path,
  $l_{mfp}=R_c/\rho_{1D}$, from the length dependent resistance
  $R(\mathcal{L},T) = R_c(T) (1+\mathcal{L}/l_{mfp}(T))$. For the SiNW we
  obtain a room-temperature mean-free-path of 3.1\,nm}\ }\BibitemShut {NoStop}%
\end{thebibliography}

\end{document}